\documentclass[aps,pra,twocolumn,floatfix]{revtex4-1}

\pdfoutput=1

\usepackage[utf8]{inputenc}
\usepackage{amssymb,amsmath}
\usepackage{physics}
\usepackage{graphicx}
\usepackage[usenames,dvipsnames]{xcolor}
\usepackage{changes}
\usepackage[colorlinks,bookmarks=false,citecolor=blue,linkcolor=cyan,urlcolor=blue]{hyperref}

\newcommand{\cl}[1]{\mathcal{#1}} 
\def\ud{\mathrm{d}}
\def\dsgamma{\gamma_\text{3D}}
\def\dsgamma{\varrho}

\newcommand{\bmx}{\begin{pmatrix}}
\newcommand{\emx}{\end{pmatrix}}
\newcommand{\bsmx}{\begin{smallmatrix}}
\newcommand{\esmx}{\end{smallmatrix}}
\newcommand{\vect}[2]{\begin{pmatrix} {#1} \\ {#2} \end{pmatrix}}

\begin{document}

\title{Nonequilibrium polariton dynamics in a Bose-Einstein condensate coupled to an optical cavity}
\author{G\'abor K\'onya}
\affiliation{Institute for Solid State Physics and Optics, Wigner Research Centre for Physics, Hungarian Academy of Sciences, H-1525 Budapest P.O. Box 49, Hungary}
\author{D\'avid Nagy}
\affiliation{Institute for Solid State Physics and Optics, Wigner Research Centre for Physics, Hungarian Academy of Sciences, H-1525 Budapest P.O. Box 49, Hungary}
\author{Gergely Szirmai}
\affiliation{Institute for Solid State Physics and Optics, Wigner Research Centre for Physics, Hungarian Academy of Sciences, H-1525 Budapest P.O. Box 49, Hungary}
\author{Peter Domokos}
\affiliation{Institute for Solid State Physics and Optics, Wigner Research Centre for Physics, Hungarian Academy of Sciences, H-1525 Budapest P.O. Box 49, Hungary}

\begin{abstract}
We study quasiparticle scattering effects on the dynamics of a homogeneous Bose-Einstein condensate of ultracold atoms  coupled to a single mode of an optical cavity. The relevant excitations, which are polariton-like mixed excitations of photonic and atomic density-wave modes, are identified. All the first-order correlation functions are presented by means of the Keldysh Green's function technique. Beyond confirming the existence of the resonant enhancement of Beliaev damping, we find a very structured spectrum of fluctuations. There is a spectral hole burning at half of the recoil frequency reflecting  the singularity of the Beliaev scattering process. The effects of the photon-loss dissipation channel and that of the Beliaev damping due to atom-atom collisions can be well separated. We show that the Beliaev process does not influence the properties of the self-organization criticality.
\end{abstract}

\maketitle

\section{Introduction}
\label{sec:intro}

In open quantum systems, an external coherent driving together with the energy dissipation into the environment can lead to interesting new features of quantum critical phenomena 
\cite{Diehl2010Dynamical,Sieberer2013Dynamical,chiacchio2018tuning,zheng2018anomalous,Cheung2018Emergent,lang2016critical,Marino2016Quantum,Hwang2018Dissipative,Soriente2018Dissipation}. Both the driving and damping have substantial effects on the spectrum of quantum fluctuations and, thereby, on the nature of dissipative phase transitions.  It has recently been shown that the critical exponent of the diverging fluctuations at a critical point is determined by the spectral properties of the relevant dissipation channel \cite{Nagy2015Nonequilibrium,Nagy2016Critical}. Recent experiments provided for measured values of the critical exponent in the dynamical phase transition of ultracold atoms coupled to the field in a high-finesse optical resonator \cite{Brennecke2013Realtime,Landig2015Measuring}. It was also shown that the decay of the coherent excitations of the Bose-Einstein condensate (BEC) had a noticeable effect on the detected quantum fluctuations. Nevertheless, this dissipation process has not yet been described within a model that would enable us to discuss its impact on the observed criticality. In this paper we present a detailed microscopic theory which takes into account all the components  of the experiment that can be relevant to the dissipation and criticality.

We consider the dissipative processes in a condensate of ultracold bosonic atoms that are coupled to a single mode of a high-finesse optical cavity. This system is known to produce a so-called self-organization phase transition which was thoroughly studied both theoretically \cite{Nagy2008Selforganization,larson2008quantum,fernandez-vidal2010quantum,keeling2010collective} and experimentally \cite{Baumann2010Dicke,Baumann2011Exploring,Schmidt2014Dynamical,Klinder2015Observation,Kollar2017Supermode,Landig2016Quantum}. It takes place at a critical value of the external laser driving strength, the \emph{control parameter}, where the initially homogeneous atom cloud illuminated from the side undergoes an ordering into a wavelength periodic pattern matching the cavity mode function. Underlying this phase transition there is a long-range interaction between the atoms mediated by the optical field in the cavity. The cavity mode geometrically selects a quasiparticle excitation of the condensate with which it forms an atom-field \emph{polariton mode}. This polariton is the soft mode of the continuous phase transition as its eigenfrequency vanishes at the critical point. This self-organization phase transition was recently extended to spin texture formation \cite{landini2018formation,kroeze2018spinor} and also to higher symmetries \cite{moodie2018generalized,leonard2017supersolid}.

The polariton is subject to damping because the cavity mode is lossy: there is a continuous photon leakage through the mirrors into the free-space modes. This dissipation channel, in the present case, is very well understood and can be treated as usual linear relaxation process with exponential decay accompanied by quantum fluctuations with white-noise spectrum.  On the other hand, the short-range interaction between the atoms, namely the s-wave atom-atom collision leads to highly non-trivial dissipative processes. One of the possible scattering processes leads to the so-called Beliaev damping \cite{beliaev1958energy,hugenholtz1959ground,graham2000langevin,VanRegemortel2017Spontaneous}. The atomic density-wave quasiparticle in the presence of the condensate decays into two other density-wave modes. In a recent paper we showed for the lossless cavity that the resulting Beliaev damping undergoes a resonant enhancement as the control parameter is varied \cite{Konya2014Photonic,Konya2014Damping}. This prediction was in qualitative agreement with experimental observations \cite{Brennecke2013Realtime}. In the theory, we treated the atom-atom interaction within Markov approximation, which led to characterizing the Beliaev damping by means of a single decay parameter. The following questions arise naturally. Can the Markovian approximation be used in such situations with enhanced dissipation? Furthermore, is it valid to treat the cavity free from losses? Can there be a nontrivial interplay between the two dissipation channels?

In the present paper we make a significant progress in describing the Beliaev damping process and answer the above questions. We adopt  the Keldysh-type Green's function approach which enables us to deal with the coupling of lossy subsystems with arbitrary spectral density function and to simplify the higher-order processes in a systematic expansion of the Dyson equation.  Owing to this general approach, we can calculate the  spectrum of quantum fluctuations of the polariton mode even if the collisional  phonon bath leads to non-Markovian dynamics and we can also include the effects of cavity photon loss.

This paper is organized as follows. Following this introduction, in Sect.~\ref{sec:model} we present the model for the laser-driven Bose-Einstein condensate strongly coupled to a single mode of a high-finesse optical cavity. In Sect.~\ref{sec:keldysh} we present the Keldysh-Green's functions for the full problem, and integrate out the quasi-particle modes of the condensate other than the one coupled to the optical mode. This calculation leads to describing the effect of phonon modes by means of a reservoir coupling density function. Thereby we prove our previous conjecture that the Beliaev process can be mapped, to leading order, on the problem of linear coupling to a reservoir composed of bosonic modes. The coupling density function characterizing the effective reservoir, however, can be a strongly patterned spectral function.  In Sect.~\ref{sec:cpdf}, we calculate this coupling function in the case of immersing the quasiparticle into a three-dimensional condensate. We determine the first-order correlation function of the quasiparticle excitations in Sect.~\ref{sec:corrfluct}, where numerical examples demonstrate the significantly non-Markovian character of the dissipative processes.

\section{Bose-Einstein condensate in a high-finesse optical cavity}
\label{sec:model}

\begin{figure}[tb!]
\centering
\includegraphics[width=\linewidth]{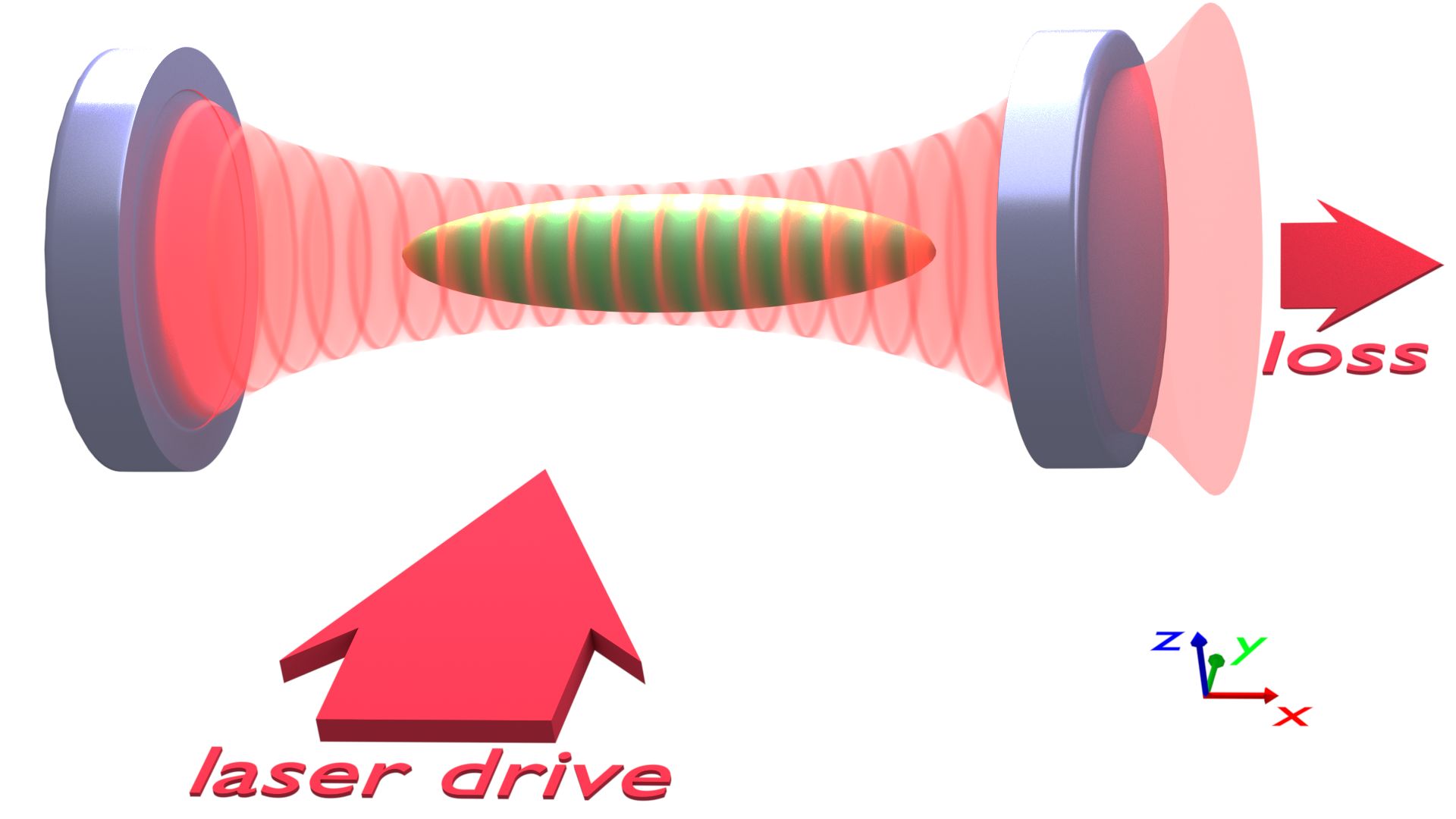}
\caption{The schematic picture of the system. An atomic ensemble inside a Fabry-Pérot cavity is pumped from the side by a laser close to resonance with the cavity. The atomic gas undergoes self-organization for strong enough laser drive: the atoms scatter photons from the laser to the cavity and may create a classical field serving as an optical lattice trapping them even further in the optimal scattering positions.}
\label{fig:scheme}
\end{figure}
We consider a trapped cloud of ultracold atoms coupled to a single-mode of a high-finesse optical resonator in the geometry corresponding to the self-organization experiments \cite{Baumann2010Dicke}. The atoms are illuminated from the side, from a direction perpendicular to the cavity axis, with a far-detuned laser standing wave. The driving field angular frequency $\omega_L$ is close to the cavity mode resonance $\omega_C$, the detuning is defined as  $\Delta_C = \omega_L - \omega_C$, whereby there is efficient photon scattering off the atoms between the quantized cavity mode and the classical laser driving field. The atoms are represented by the quantized matter-wave field $\hat{\Psi}(x)$, whereas the single cavity mode with spatial mode function $\cos(k\, x)$ is described by the annihilation and creation operators $a, a^\dagger$. The total grand canonical Hamiltonian reads then, in units of $\hbar$,  
\begin{multline}
\label{eq:ham1}
\hat{K} = \hat{H} - \mu \, \hat{N} \\ 
= -\Delta_{C} \, \hat{a}^\dag \, \hat{a} \, 
+ \, \int_{-\frac{L}{2}}^{+\frac{L}{2}} dx \; \hat{\Psi}^\dag (x) 
\left( -\frac{1}{2m} \frac{d^2}{dx^2} - \mu \right) \hat{\Psi}(x)  \\
+ \frac{g}{2} \, \int_{-\frac{L}{2}}^{+\frac{L}{2}} dx \; \hat{\Psi}^\dag (x) \, \hat{\Psi}^\dag (x) \, 
\hat{\Psi} (x) \, \hat{\Psi} (x) \; ,\\
+ \eta \left( \hat{a}^\dag + \hat{a} \right) \, 
\int_{-\frac{L}{2}}^{+\frac{L}{2}} dx \; \hat{\Psi}^\dag (x) \, \cos(k \, x) \, \hat{\Psi} (x) \; ,
\end{multline}
where $\hat{H}$ is the Hamiltonian, $\hat{N}$ is the particle number operator and $\mu$ is the chemical potential. The first term is the energy of the photon field in the frame rotating with the driving laser frequency $\omega_L$. Then the atomic energy term follows, which includes the kinetic energy for mass $m$, and the short-range contact interaction  with effective coupling strength $g$. The last term describes the interaction that arises from the photon scattering: it amounts effectively to a driving of the cavity mode with amplitude $\eta$ incorporating atomic and laser parameters, such as the atomic dipole strength, the detuning from the atomic resonance and the laser field strength.  This $\eta$ can experimentally be the control parameter of the system.

For notational simplicity, we use only the spatial coordinate $x$, however, all the expressions can be trivially generalized to three dimensions, considering simply $x$ as a three-component vector. Since the trapping potential geometry is irrelevant, as long as its length scale is much larger then the optical wavelength scale, we introduce a fictitious box containing the atom cloud with length $L$. It will be taken in the limit $L\rightarrow \infty$. The atomic field can be expanded in plane-wave basis
\begin{equation}
\hat{\Psi} (x) = \frac{1}{\sqrt{L}} \sum_{p \, \in \, \cl{P}^{(L)}} \hat{h}_p \; e^{i \, p \, x} \; ,
\end{equation}
where the bosonic annihilation operator $\hat{h}_p$ is associated with the mode with wavenumber $p$, element of the index set $\cl{P}^{(L)} = \left\{ \, \frac{2\pi}{L} \cdot j \,  |  \, j \, \in \, \mathbb{Z} \, \right\}$.

Throughout the paper, we assume that the driving strength is smaller than the critical value of the self-organization phase transition. It means that the cavity contains no coherent mean-field of photons and there is no optical dipole potential modulating the spatial distribution of the condensate. We can then safely assume that the atomic gas is prepared as a Bose-Einstein condensate in the lowest momentum state $p=0$, and that the population of thermal excitations are very small. Thus, we can perform the replacement $\hat{h}_0, \hat{h}^\dagger_0 \rightarrow \sqrt{N_c}$, where $N_c$ is the number of atoms in the Bose-condensed cloud. As $N_c$ is much larger than the occupation of all the other single-particle modes, the grand canonical Hamiltonian can be rearranged according to the order of the product of the ladder operators ($\hat{a},\hat{a}^\dagger,\hat{h}_p,\hat{h}^\dagger_p$ with $p\neq0$) such that,
\begin{equation}
\hat{K} = \hat{K}_0 +\hat{K}_1 + \hat{K}_2 + \hat{K}_3 + \hat{K}_4 \; ,
\end{equation}
The zeroth order term $\hat{K}_0$ is an uninteresting scalar shift of the energy and can be dropped. The next order, $\hat{K}_1$, contains all the terms with only one ladder operator. This term is generally used to determine the mean-field condensate density and wavefunction by setting $\hat{K}_1=0$. In our simple case of a homogeneous condensate, this term is automatically zero due to momentum conservation. Finally, we neglect the smallest 4th order term which comprises only single-particle operators. We keep the second and third orders, the former being a Bogoliubov-type quadratic Hamiltonian; the latter contains the scattering of excitations over the highly-populated BEC wavefunction. These scattering processes are responsible for the damping of excitations and are in the focus of this paper.

In the thermodynamic limit, where $N_c \rightarrow \infty$, $L \rightarrow \infty$ such that $N_c/L$ is constant, each term of the Hamiltonian must be extensive, i.e., proportional to $N_c$ or $L$. Then, we must assume that all ladder operators are asymptotically in the order of $\sqrt{L}$ and we need to introduce the coupling constant $\tilde{g} = g \, {N_c}/{L}$ and $y=\sqrt{2\, N_c} \eta$ such that they have a finite value in the thermodynamic limit.

According to the last line of Eq. \eqref{eq:ham1}, momentum conservation ensures that the cavity directly couples only atoms with momentum $\pm k$ to the condensate. As higher momentum modes play little role in the self-organization transition \cite{konya2011multimode}, we neglect them. Trigonometric identities restrict this coupling further, eventually to a single atomic state. Instead of the $\pm k$ plane wave modes, we introduce a cosine and a sine mode, and their ladder operators accordingly,
\begin{subequations}
\begin{align}
\hat{h}^\dag_{\pm k} &= \frac{1}{\sqrt{2}} 
\left( \, \hat{c}^\dag \pm i \, \hat{s}^\dag \, \right) \; , \\ 
\hat{h}_{\pm k} &= \frac{1}{\sqrt{2}} 
\left( \, \hat{c} \mp i \, \hat{s} \, \right) \; .
\end{align}
\end{subequations}
The newly introduced operator $\hat{c}$ is the annihilation operator of the cosine mode, while $\hat{s}$ is that of the sine mode. The cosine-density wave and the cavity mode form a hybrid excitation with both atomic and photonic characters and will be referred to as a \emph{polariton}.

\begin{figure}
\includegraphics[width=0.99\linewidth]{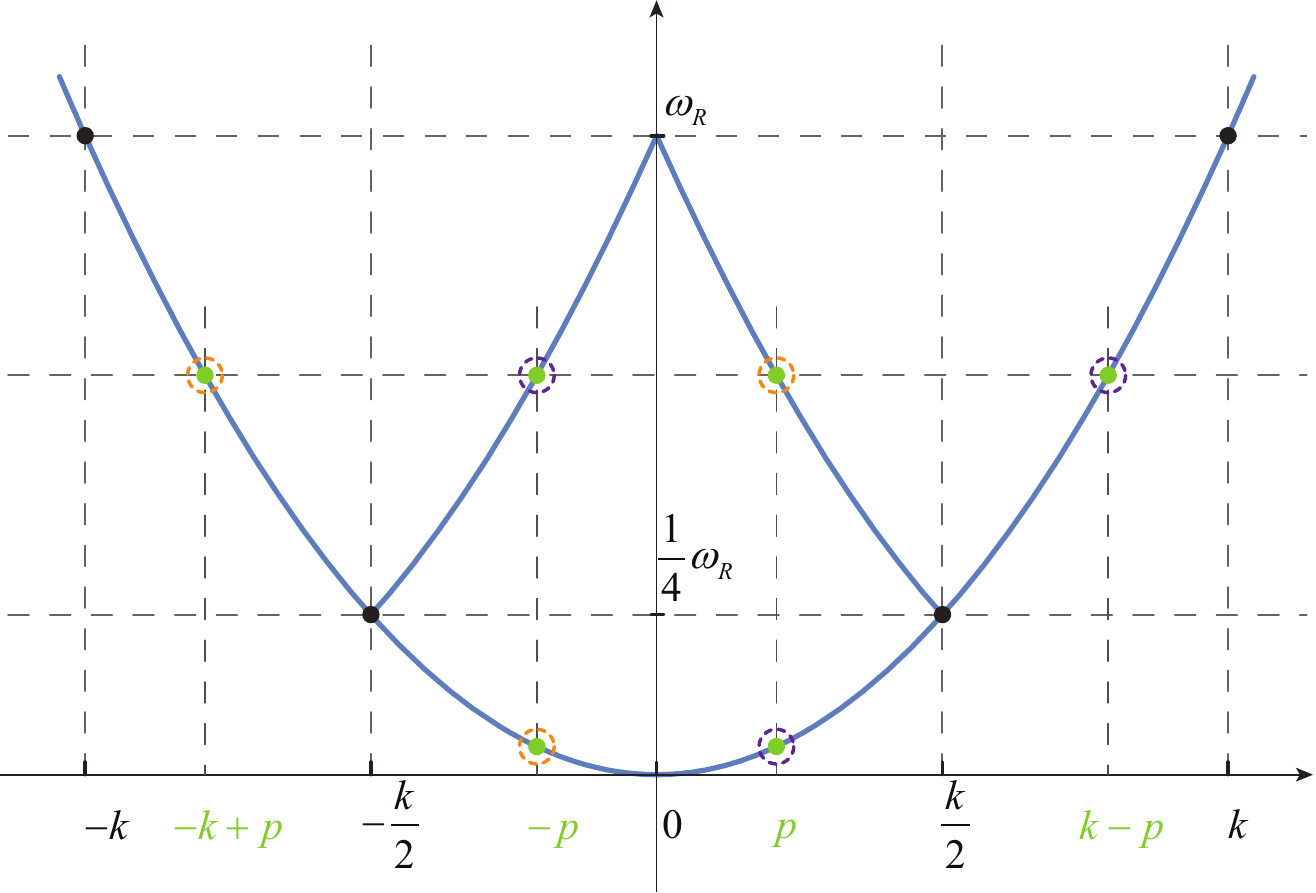}
\caption{Dispersion of single-particle excitations of the condensate forming a bath for the density-wave mode selected by the optical resonator mode. The first band expands in the range $(0,\omega_R/4)$, the second band is  $(\omega_R/4,\omega_R)$. Higher bands are not plotted but are included in the calculation.}
\label{fig:parabola}
\end{figure}

We introduce further simplifications that can be invoked in the subspace of the polariton dynamics. These are the following:
\begin{itemize}
\item We will see later that the low momentum $p\approx 0$ part of the spectrum is irrelevant for the dynamics of the polariton excitation having $\pm k$ momentum. Therefore we can safely neglect the second-order collisional Bogoliubov terms $\hat{h}_p \hat{h}_{-p}$, $\hat{h}_p^\dagger \hat{h}_{-p}^\dagger$ and $\hat{h}_p^\dagger \hat{h}_p$ from $\hat{K}_2$ and take $\mu \simeq 0$ for the sake of consistency. This approximation amounts to replacing the linear part of the Bogoliubov spectrum near $p\approx 0$ by the quadratic dispersion relation valid for higher quasi-momentum excitations.  
\item  In Eq.~\eqref{eq:ham1} only the cosine mode survives the integration in the last line from the laser pump. In the second line, we neglect the collisional terms of the odd-parity $\hat s$ mode.
\item In the third-order term $\hat{K}_3$, we keep only terms involving a polariton operator $\hat{c}$ or $\hat{c}^\dagger$. The dominant term is the one originating from s-wave scattering with a condensate particle. The remaining two atomic operators must have momentum $p$ and $k-p$ in order to satisfy momentum conservation.
\end{itemize}
With these approximations we arrive at
\begin{multline}
 \label{eq:K_final}
\hat{K} = -\Delta_{C} \, \hat{a}^\dag \, \hat{a} \, + \, \omega_R \, \hat{c}^\dag \, \hat{c} \, 
+ \, \frac{1}{2} \, y \, \left( \hat{a}^\dag + \hat{a} \right) \, \left( \hat{c}^\dag + \hat{c} \right) \\
+ \, \sum_{p } \omega_p \, \hat{h}_p^\dag \, \hat{h}_p 
+ \frac{2}{\sqrt{2 \, N_c}} \, \tilde{g} \\
\times\, \sum_{p<{k}/{2}}  \left(  \hat{c}^\dag \, \hat{h}_{p} \, \hat{h}_{k-p} 
+ \hat{c}^\dag \, \hat{h}_{-p} \, \hat{h}_{-k+p} + \text{h.~c.} \right)  \; ,
\end{multline}
where we use the recoil frequency $\omega_R=k^2 / 2m$ and the approximate dispersion relation $\omega_{p} = {p^2}/{2m}$ illustrated in Fig.~\ref{fig:parabola}. We often refer to the modes $h_{\pm p}$ as \emph{phonons}.The summation goes over the set $p\in \cl{P}^{(L)}  \setminus \{ 0 , +k,  -k \}$.
Each term  $\hat{h}_{p} \, \hat{h}_{k-p}$ in the summation has the symmetry $p \leftrightarrow k-p$. This way, the summation for the atomic single-particle momentum states is confined to the $0< p < k/2$ interval. The negative side of the dispersion is taken into account by a factor of 2 in front of the sum.

\section{Keldysh approach to the dissipative system}
\label{sec:keldysh}

The Hamiltonian in Eq.~\eqref{eq:K_final} describes two coupled boson modes, $\hat a$ and $\hat c$, the latter one interacting also with many phonon modes $\hat h_p$. In the limit of large length $L$, these phonon modes form a continuum that can be considered as a bath acting on mode $\hat c$. We describe this driven-dissipative quantum system with the help of the Keldysh path integral technique, the details of which can be found, e.g., in Refs. \cite{kamenev2011field,rammer2007quantum,sieberer2016keldysh,dallatorre2013keldysh}. 

The key quantities of our investigation are the interacting Green's functions of the cosine mode $\hat c$, and the photon mode $\hat a$. The Green's functions of mode $\hat c$ are defined as
\begin{multline}
\label{eq:intgrmat}
\mathbf{G}_c(t,t') \equiv \begin{pmatrix}
G^K_c(t,t') & G^R_c(t,t') \\
G^A_c(t,t') & 0
\end{pmatrix}
\\=
(-i)\begin{pmatrix}
\langle c^{cl} (t) \, c^{cl \, *} (t')\rangle & \langle c^{cl} (t) \, c^{q \, *} (t')\rangle \\
\langle c^{q} (t) \, c^{cl \, *} (t')\rangle & 0
\end{pmatrix}.
\end{multline}
Inside the averages, the variables are weighted with the Keldysh action and are integrated over in a path integral \cite{sieberer2016keldysh}. The Green's functions of the photon mode are defined in a completely similar way.
We split the Keldysh action to three non-interacting components and two interaction parts,
\begin{equation}
\label{eq:Keldactf}
S = S_{a} + S_{c} + S_{h} + S_{ch} + S_{ac}\; .
\end{equation}
After introducing classical and quantum variables for the ladder operators, the non-interactiong Keldysh action for modes $\hat a$, $\hat{c}$ and for the phonons, read as
\begin{equation}
\label{eq:Sa}
S_{a} = \int_{-\infty}^{+\infty} dt \, 
\begin{pmatrix} 
a^{cl \, *} (t) & a^{q \, *} (t)
\end{pmatrix} 
\, \mathbf{G}_a^{(0)\, -1} \,
\begin{pmatrix}
a^{cl} (t) \\
a^{q} (t) 
\end{pmatrix} \; ,
\end{equation}
\begin{equation}
\label{eq:Sc}
S_{c} = \int_{-\infty}^{+\infty} dt \, 
\begin{pmatrix} 
c^{cl \, *} (t) & c^{q \, *} (t)
\end{pmatrix} 
\, \mathbf{G}_c^{(0)\, -1} \,
\begin{pmatrix}
c^{cl} (t) \\
c^{q} (t) 
\end{pmatrix} \; ,
\end{equation}
and
\begin{equation}
\label{eq:Sh}
S_{h} = \sum_{p} \, \int_{-\infty}^{+\infty} dt \, 
\begin{pmatrix} 
h_{p}^{cl \, *} (t) & h_{p}^{q \, *} (t)
\end{pmatrix} 
\, \mathbf{G}_p^{(0)\, -1} \, 
\begin{pmatrix}
h_{p}^{cl} (t) \\ 
h_{p}^{q} (t)
\end{pmatrix} \; ,
\end{equation}
with the inverse free Green's function matrices
\begin{equation}
\mathbf{G}_a^{(0) \,-1} = \begin{pmatrix}
0 & i \partial_t + \Delta_C - i \, \kappa \\
i \partial_t + \Delta_C + i \, \kappa& 2 \, i \, \kappa
\end{pmatrix},
\end{equation}
\begin{equation}
\label{eq:nonintGc}
\mathbf{G}_c^{(0) \,-1} = \begin{pmatrix}
0 & i \partial_t - \omega_R - i \, \epsilon \\
i \partial_t - \omega_R + i \, \epsilon & 2 \, i \, \epsilon
\end{pmatrix},
\end{equation}
and
\begin{equation}
\mathbf{G}_p^{(0)\, -1} =
\begin{pmatrix}
0 & i \partial_t - \omega_{p} -i \gamma_{p}\\
i \partial_t - \omega_{p} +i \gamma_{p} & 2 \, i \, \gamma_{p} ( 2 \, \bar{n}_{p} + 1 )
\end{pmatrix}\; ,
\end{equation}
respectively. The photon mode $\hat a$ has a high frequency and $\Delta_C$ is referenced to the driving frequency $\omega_L$. The decay of this mode is unaffected by the interaction with the mode $\hat c$. Moreover, the flat reservoir spectrum at high frequencies ensures the validity of a Markovian approximation, which is reflected by using a single constant parameter $\kappa$, half of the photon loss rate, in the Keldysh component. 
The parameter $\epsilon$ is an infinitesimal regularization parameter. Its actual value is irrelevant, since the Green's function for mode $\hat c$ is regularized due to the interaction with the bath of phonon modes. The phonon modes themselves are also decaying, although this mechanism originates from the neglected $K_3$ terms describing atom-atom collisions of three phonon operators, i.e., not including any polariton mode. Higher order terms or other physical processes may also contribute to the phonon decay. Rather than modelling these processes at a microscopic level, here we simply introduce phenomenologically a linewidth $\gamma_p$ for the phonon modes. We take into account finite temperature via the thermal population of the phonon modes $\bar{n}_{p}=(e^{\beta \omega_p}-1)^{-1}$, with $\beta=1/k_BT$ the inverse temperature.

Finally, the interaction terms are 
\begin{multline}
S_{ch} = - \frac{\tilde g}{\sqrt{N_c}} \, \int_{-\infty}^{+\infty} dt \, \sum_p 
\Biggl( 
c^{cl \, *} \, h_{p}^{cl} \, h_{k-p}^{q} + 
 c^{cl \, *} \, h_{p}^{q} \, h_{k-p}^{cl} \\
+ c^{q \, *} \, h_{p}^{cl} \, h_{k-p}^{cl}
+ c^{q \, *} \, h_{p}^{q} \, h_{k-p}^{q} + c.~c.
\Biggr) \; ,
\label{eq:sint}
\end{multline}
describing the interaction between the cosine mode and the phonons, while
\begin{multline}
\label{eq:Sac}
S_{ac} = -\frac{y}{2}\int_{-\infty}^{+\infty} dt \, 
\Big[(a^{q} + a^{q \, *})(c^{cl} + c^{cl \, *}) \\
+ (a^{cl} + a^{cl \, *})(c^q + c^{q \, *})\Big]\,
\end{multline}
represents the interaction between the cosine mode and the cavity photons.

\begin{figure}[!tb]
\centering
\includegraphics{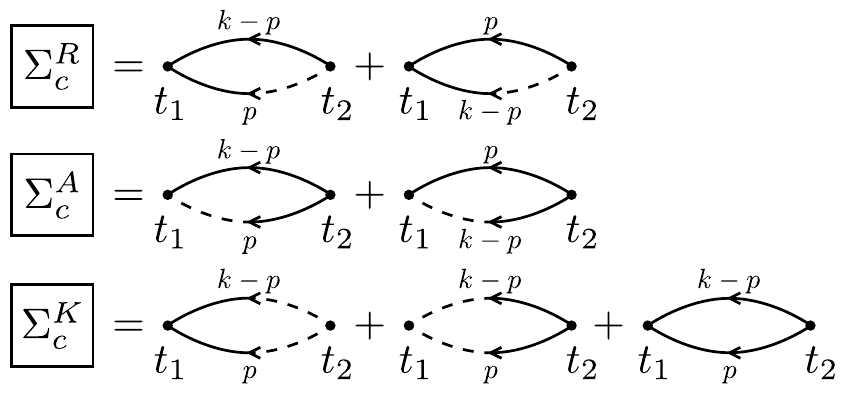}
\caption{The self energies of the cosine-mode Green's functions. The solid lines sand for the classical components of the phonon fields, while the dashed lines represent their quantum parts.}
\label{fig:self_energ}
\end{figure}

In order to arrive to the effective dynamics of the two relevant degrees of freedom, namely $\hat a$ and $\hat c$, we integrate out the $\hat h_p$ modes in one-loop level. As a result of the integration, a new decay channel to the mode $\hat c$ emerges from the phonon bath. We proceed by calculating an intermediate Green's function for $\hat c$ in perturbation theory with Eqs.~\eqref{eq:Sc} and \eqref{eq:Sh} as the free system and Eq.~\eqref{eq:sint} as the perturbation. The intermediate Green's function is expressed with the Dyson equation (after Fourier transformation),
\begin{equation}
\widetilde{\boldsymbol{G}}_c^{-1} (\omega) = \boldsymbol{G}_c^{(0) \, -1} (\omega) - \boldsymbol{\Sigma}_c (\omega) \;  ,
\label{eq:internalG}
\end{equation}
where we introduced the self energy $\boldsymbol{\Sigma}_c(\omega)$,
the contribution of the irreducible graphs connecting to two external
points \cite{rammer2007quantum}. The above intermediate Green's
function defines the effective action $\widetilde{S}_c$ in a similar manner as the noninteracting Green's function~\eqref{eq:nonintGc} defines the bare action \eqref{eq:Sc}. Now, the renormalized action $\widetilde{S}_c$ includes also the decay of mode $\hat c$ into the phonon bath. In
Appendix~\ref{app:pertkel}, we outline the main steps leading to the
Dyson equation for mode $\hat c$. The corresponding Feynman diagrams
of the cosine mode self energies are depicted in
Fig~\ref{fig:self_energ}. Their contribution is
\begin{subequations}
 \label{eq:nonlinear_self}
\begin{equation}
\Sigma^{R/A}_c (\omega) = \frac{2 \tilde{g}^2}{N_c}  \, \sum_{p<k/2} \frac{ \bar{n}_{p} + \bar{n}_{k-p} + 1 }{\left( \omega - \omega_{p} - \omega_{k-p} \right) 
\pm i \left( \gamma_{p} + \gamma_{k-p} \right) } \; ,
\end{equation}
where the upper sign stands for the retarded, and the lower sign for the advanced part of the self energy. The Keldysh component is evaluated to be
\begin{multline}
\label{eq:SK1}
\Sigma^{K}_c (\omega) = -i \,\frac{4 \tilde{g}^2}{N_c}  \, \sum_{p<k/2}
\left( 2 \, \bar{n}_{p} \, \bar{n}_{k-p} + \bar{n}_{p} 
+ \bar{n}_{k-p} + 1 \right) \\
\times \frac{\gamma_{p} + \gamma_{k-p}}{\left(\omega-\omega_{p}-\omega_{k-p} \right)^2 + \left( \gamma_{p} + \gamma_{k-p} \right)^2} \; .
\end{multline}
\end{subequations}

By introducing a fictitious Beliaev quasiparticle with 
\begin{subequations}
\label{eqs:definitions}
\begin{align}
\omega^{(B)}_{p} &= \omega_p +\omega_{k-p},\\
\gamma^{(B)}_{p} &= \gamma_p +\gamma_{k-p},\\
g_p^2 &= 2 \tilde{g}^2 (\bar{n}_{p} + \bar{n}_{k-p} + 1)\; ,\label{eq:effcc}
\end{align}
\end{subequations}
the self energies take the following form:
\begin{subequations}
 \label{eq:linear_self}
\begin{equation}
\Sigma^{R/A}_c (\omega) = \frac{1}{N_c} \sum_{p<k/2} \,  
\frac{g^{2}_p}{\omega - \omega^{(B)}_{p} \pm i \gamma^{(B)}_{p} } \; ,
\end{equation}
\begin{multline}
\label{eq:SK2}
\Sigma^{K}_c (\omega) = -2i \, \frac{1}{N_c} \,  \sum_{p<k/2} g^{2}_p \,
\left( 2 \, \bar{n}^{(B)}_{p} + 1 \right) \\ \times\frac{\gamma^{(B)}_{p}}{\left(\omega-\omega^{(B)}_{p} \right)^2 + \gamma^{(B) \, 2}_{p} } \; ,
\end{multline}
\end{subequations}
which are the self energies corresponding to a bath which is linearly coupled to the mode $\hat c$. For consistency, the newly introduced occupation number must be 
\begin{equation}
\label{eq:newocup}
\bar{n}_{p}^{(B)} = \frac{1}{e^{\beta \, \left( \omega_{p} + \omega_{k-p} \right) } -1}\ .
\end{equation}
The numerators of Eq.~\eqref{eq:SK1} and Eq.~\eqref{eq:SK2} are, in fact, the same, since
\begin{multline*}
 1+  \frac{1}{\bar{n}_{p}^{(B)}}  = e^{\beta \, \left( \omega_{p} + \omega_{k-p} \right)} \\ = e^{\beta\omega_{p}} e^{\beta \omega_{k-p}} = \left( 1+  \frac{1}{\bar{n}_{p}} \right)\, \left( 1+  \frac{1}{\bar{n}_{k-p}}\right)\,.
\end{multline*}

\section{The coupling density function of the phonon bath}
\label{sec:cpdf}

Once the interaction with the phonons is expressed in the form of a linear coupling to a bath, all the properties of the self-energy functions can be originated from a spectral function, the so-called \emph{coupling-density function}. It is defined by
\begin{equation}
\label{eq:cdf}
\rho(\omega) = \frac{1}{N_c} \, \sum_{p < k/2} {g}_p^2\,  
\delta \left( \omega - \omega^{(B)}_{p} \right) \; .
\end{equation}
In the following, first we calculate this coupling-density function from the microscopic model, and then we explicitly give the self energies of the polariton excitation, 
\begin{subequations}
 \label{eq:self_energies}
\begin{equation}
\Sigma^{R/A}_c (\omega) = \int_{-\infty}^{+\infty} \, d \omega' \, \rho(\omega') \,
\frac{1}{\omega - \omega' \pm i \gamma^{(B)} } \; ,
\end{equation}
\begin{equation}
\Sigma^{K}_c (\omega) = -2i \, \int_{-\infty}^{+\infty} \, d \omega' \, \rho(\omega')
\cdot \frac{\gamma^{(B)}}{\left(\omega-\omega' \right)^2 + \gamma^{(B) \, 2} } \; .
\end{equation}
\end{subequations}

We assume that the damping rate of the phonons $\gamma^{(B)}_p$ is
small for all the relevant phonon modes that participate in the
Beliaev process, thus, taking the limit $\gamma^{(B)}_p \rightarrow
0$, the self energies become
\begin{subequations}
 \label{eq:self_energies2}
\begin{eqnarray}
  \Sigma^{R/A}_c (\omega) & = & {\cal P}\int_0^\infty\frac{\rho(\omega^{'})}{\omega-\omega^{'}}
 \,\ud\omega^{'} \mp i\pi\rho(\omega) \;, \\
\Sigma^{K}_c (\omega) & = & 2\pi{}i\rho(\omega) \;.
\end{eqnarray}
\end{subequations}
We write the Keldysh component for zero temperature ($T=0$).
With this, the collisional phonon interaction is incorporated into these self-energies. This can be then the starting point to study the dynamical behavior of the polariton excitation.

In order to get a numerical estimate, we consider a ${}^{87}$Rb condensate of $N_c=10^5$ atoms at $T=0$, in a 3-dimensional harmonic trap of mean frequency $\bar\omega=2\pi \times 142$ Hz, which is well in the Thomas-Fermi limit. The scattering length is $a=5.29$ nm, the wavelength is $\lambda=780$ nm, and the mass is $m=1.443\times 10^{-25}$ kg.
At zero temperature, the effective coupling constant, Eq.~\eqref{eq:effcc}, simplifies to $g_p=\sqrt{2}\tilde{g}$. In a shallow trap, the dispersion relation of the effective Beliaev quasiparticle can be approximated as 
\begin{equation}
\omega^{(B)}_{p_x \, , \, \vb{p}_{\bot}} = \frac{p_x^2}{2m} + \frac{(k-p_x)^2}{2m} + 2 \cdot \frac{\vb{p}_{\bot}^2}{2m} \; ,
\end{equation}
where $\vb{p}_{\bot}$ is the momentum orthogonal to the cavity axis direction $x$ and the wavenumber $k$.
The summation in the thermodynamic limit goes into
\begin{equation}
\sum_{p_x < {k}/{2} } \sum_{\vb{p}_{\bot}} \left( \, \ldots \, \right) \; 
\rightarrow \; \frac{V}{(2 \, \pi)^3} \int_{-\infty}^{\frac{k}{2}} \, d p_x 
\int_{\mathbb{R}^2} \, d^2 p_{\bot} \, \left( \, \ldots \, \right) \; .
\end{equation}

A straightforward evaluation of Eq.~\eqref{eq:cdf} yields 
\begin{align}
 \label{eq:rho_3D}
\rho(\omega) = & \varrho \; 
\sqrt{\frac{\omega}{\omega_R} - \frac{1}{2} } \; \; 
\theta \left( \frac{\omega}{\omega_R} - \frac{1}{2} \right)\; ,\\
\varrho = & \sqrt{\frac{\omega_R \, m^3}{\hbar^3}} \; \frac{g^2}{2 \, \pi^2} \; \frac{N_c}{V} \, .
\end{align}
This expression for the coupling density function is the central result of the paper. In the following, we use only this formula in order to calculate measurable correlation functions that describe the dynamical properties of the coupled photon-quasiparticle system. Using the bare s-wave scattering constant $g=4\pi\hbar a /m$ where $a$ is the s-wave scattering length, one gets for the coefficient
\begin{equation}
\frac{\varrho}{\omega_R} = \frac{4 \, \sqrt{2}}{\pi} \, \frac{N_c}{V} \, a^2 \, \lambda \; .
\end{equation}
With the above mentioned values of the parameters, the formula yields $\varrho \approx 1.4\times 10^{-2} \omega_R$. This $\varrho$ can be tuned by applying an external magnetic field to increase the scattering length.

\section{Correlations and fluctuations}
\label{sec:corrfluct}

\begin{figure*}
\includegraphics[width=0.95\linewidth]{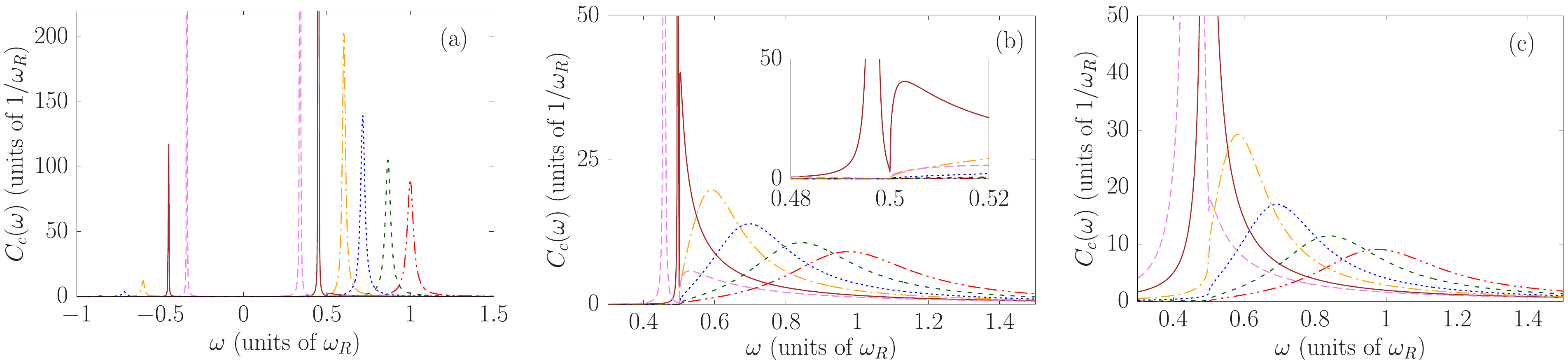}
\caption{Correlation function of the quasiparticle excitation. Large
  detuning $\Delta_C=100$ implies little mixing of the condensate
  quasiparticle with the photon mode. (a) Weak coupling to the
  environment, $\kappa=0.1$, $\dsgamma=0.01$. (b) Enhanced
  Beliaev scattering process $\dsgamma=0.1$, $\kappa=0.1$. (c)
  Enhanced photon loss rate $\kappa=10$, $\dsgamma=0.1$.  (All
  angular frequencies are expressed in units of $\omega_R$.)  The
  correlation function is plotted for different coupling strengths $y$
  approaching the critical value $y_c$: $y=0$ (dashed-double-dotted red), $y/y_c=
  0.5$ (short-dashed green), $0.7$ (dotted blue), $0.8$ (dashed-dotted
  orange), $0.9$ (solid brown), $0.95$ (long-dashed
  magenta). }
\label{fig:Correlation_c_mode}
\end{figure*}

After integrating out the phonon bath, we take into account its effect
solely by means of the spectral function $\rho(\omega)$, Eq.~(\ref{eq:rho_3D}).
The total Keldysh action, Eq.~(\ref{eq:Keldactf}), reduces to
\begin{equation}
  \label{eq:S_twomode}
  S = S_a + \widetilde{S}_c + S_{ac}\,,
\end{equation}
that corresponds to the problem of two interacting bosonic modes which are coupled to
their respective environments. In Fourier space,
\begin{equation}
\label{eq:Safour}
S_a = \int \frac{\ud\omega}{2\pi}\left(a^{\rm cl *}, a^{\rm
  q *}\right) \bmx 0 & \omega+\Delta_C-i\kappa
\\ \omega+\Delta_C + i\kappa & 2i\kappa \emx \vect{a^{\rm
    cl}}{a^{\rm q}}
\end{equation}
describes the cavity mode decaying to a ``flat'' reservoir with a rate $\kappa$, and
\begin{equation}
\label{eq:Seff_c}
\widetilde{S}_c = 
\int\frac{\ud\omega}{2\pi} 
\left(c^{\rm cl *}, c^{\rm q *}\right)
\bmx
 0 & \omega-\omega_R - \Sigma^{A}_c \\
\omega-\omega_R - \Sigma^{R}_c & -\Sigma^{K}_c
\emx
\vect{c^{\rm cl}}{c^{\rm q}}
\end{equation}
describes the dynamics of the phonon-damped polariton mode. Its
Beliaev-type decay to the phonon bath is incorporated in the self
energies in Eqs.~(\ref{eq:self_energies2}a,b).

In the following, we study the interplay of the phonon damped
polariton mode, and the intrinsically damped, leaky cavity mode. Since
their interaction, described by $S_{ac}$ in Eq.~(\ref{eq:Sac}),
contains counter-rotating terms, e.g., $a^{cl}c^q$, $a^{q \, *} c^{cl
  \, *}$, the variable space must be doubled by introducing fields
with negative frequencies so that the total Keldysh action,
Eq.~(\ref{eq:S_twomode}) can be expressed in a closed quadratic form
\cite{dallatorre2013keldysh}. There are classical and quantum variables for both modes, and with $\omega$ and
$-\omega$ arguments, i.e., altogether eight variables. The full problem can be expressed in a $8\times8$
matrix form of the action. On inverting the matrix, one gets access to
all the Green's functions.  The calculation is analogous to the one in
Ref.~\cite{Nagy2016Critical}. Here we can simply adopt the results for
the various Keldysh Green's function components of the two-mode
system, and insert the specific function $\rho(\omega)$ calculated in
the previous section (see Eq.~(\ref{eq:rho_3D})).
The spectrum of fluctuations are of our main interest, since these quantities are directly measurable, as it was demonstrated in recent experiments \cite{Brennecke2013Realtime}. These spectra correspond to the Fourier-transform of the correlation functions,
\begin{equation}
 C_{a}(t) = \langle \{a(t),a^\dagger(0)\}\rangle \,, \; C_{c}(t) = \langle \{c(t),c^\dagger(0)\}\rangle\,,
\end{equation}
which are given by the appropriate components of the Keldysh Green's function. 

By increasing the strength of the pumping laser, the coupling between
the cosine density wave of the condensate and the photons of the
cavity becomes stronger. As a result, a mode softening takes place,
and the polariton frequency goes down to zero at a critical point,
where the system goes through a phase transition into a superradiant
phase \cite{Nagy2008Selforganization,Nagy2010Dicke,Baumann2010Dicke}.
The critical point is at $y_c = \sqrt{(\Delta_C^2+\kappa^2) \omega_R
/ |\Delta_C|}$. The effect of a non-Markovian reservoir on the
critical behaviour has been recently studied
\cite{Nagy2015Nonequilibrium,Nagy2016Critical} by assuming a
phenomenologically defined coupling density function $\rho(\omega)$.
In the present paper, we derived analytically $\rho(\omega)$, based
on an entirely microscopic approach. We found that the coupling
density function, c.f.~Eq.~(\ref{eq:rho_3D}), does not have the kind
of power law dependence at zero frequency, with exponent $0<s<1$,
which was shown to be the necessary condition for getting a critical
exponent below 1. It is in agreement with earlier theoretical
prediction \cite{Nagy2011Critical} and experimental verification
\cite{Brennecke2013Realtime}. The Beliaev damping process of the
polariton, caused by the $\hat K_3$ term, vanishes identically below
a threshold frequency, as the conservation laws for energy and
momentum can not be fulfilled simultaneously below this threshold.
Thus, Beliaev damping has no substantial impact on the properties of
criticality, as the polariton frequency goes below this threshold
much before the critical region. In the following, we focus on the
observable consequences of the Beliaev scattering process in the
spectrum of fluctuations outside the critical region.

\begin{figure*}
\includegraphics[width=0.95\linewidth]{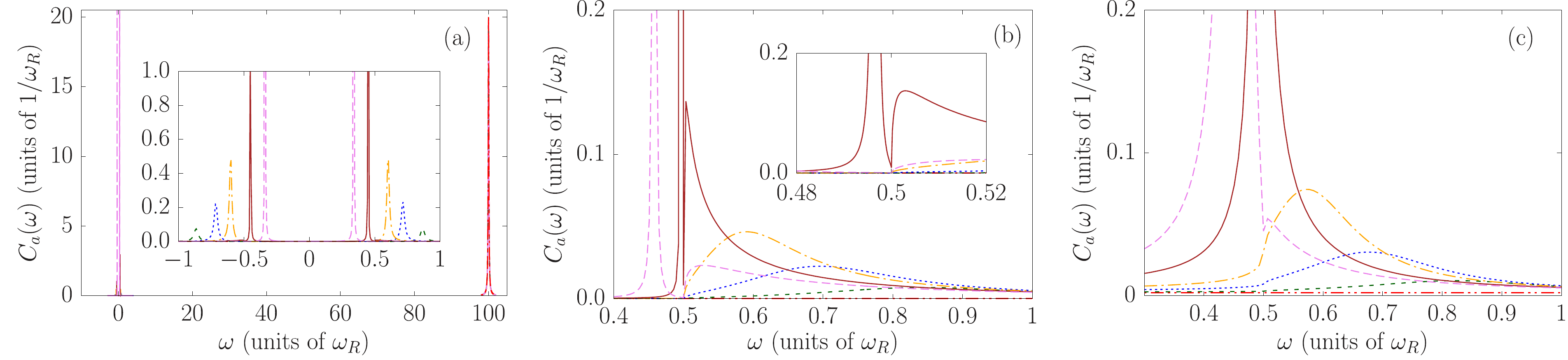}
\caption{Correlation function of the photon mode is plotted for
  different coupling strengths $y$ approaching the critical value
  $y_c$: $y=0$ (dashed-double-dotted red), $y/y_c= 0.5$ (short-dashed green), $0.7$
  (dotted blue), $0.8$ (dashed-dotted orange), $0.9$
  (solid brown), $0.95$ (long-dashed magenta).  The
  cavity resonance peak at $\Delta_C = 100$ is independent of the
  pumping strength $y$ and is shown only for weak environmental
  couplings (a); $\kappa=0.1$, $\dsgamma=0.01$. The inset
  displays the peaks around zero frequency.  (b) Increased Beliaev
  scattering broadens the peaks for $\omega > 0.5$;
  $\dsgamma=0.1$, $\kappa=0.1$. (c) Enhanced photon loss rate
  $\kappa=10$ ($\dsgamma=0.1$) broadens the peaks also for
  $\omega < 0.5$. Parameters are measured in units of $\omega_R$, and
  are the same as in Fig.~\ref{fig:Correlation_c_mode}.}
\label{fig:Correlation_a_mode}
\end{figure*}

In Fig.~\ref{fig:Correlation_c_mode} we present correlation functions of the
mode $\hat c$ for various parameter settings.  When $\omega_R < |\Delta_C|$,
the lower lying polariton mode, i.e.~the soft mode, is dominantly the
condensate quasiparticle $\hat c$. Figure \ref{fig:Correlation_c_mode}a shows
the case of weak environmental effects, that is both $\kappa$ and $\dsgamma$
are smaller than the coupling $y$. Because of the small coupling to the
environment, the spectrum manifests sharp resonance peaks corresponding to the
polariton mode frequency. The plot shows that this eigenfrequency approaches
the origin as the coupling $y$ is increased, in accordance with the mode
softening. Note that the polariton resonance appears at both the positive and
negative frequency sides, that is the result of the counter-rotating terms in
the Dicke-type light-matter interaction $S_{ac}$. We show the negative
frequency peaks only in Fig.\ref{fig:Correlation_c_mode}a and
Fig.~\ref{fig:Correlation_a_mode}a, as they behave similarly to their
counterparts at positive frequency. These 2 plots are good reference points as
they contain insignificant Beliaev damping and also pretty small photon loss
compared to other frequency scales.

In a next step, we study the variation of the shape of the spectral
lines as the decay parameters are enhanced. First, the Beliaev process
is enhanced by an order of magnitude, i.e.~$\dsgamma =0.1 \omega_R$.
In Fig.~\ref{fig:Correlation_c_mode}b the dashed-double-dotted red
line presents a simple broad peak (for $y=0$), whose width is
determined by $\dsgamma$. This spectrum has the important feature that
it strictly vanishes for $\omega < \omega_R/2$, since the
coupling-density function $\rho(\omega)$ vanishes below $\omega =
\omega_R/2$. On increasing the coupling strength $y$, the peak moves
towards lower frequencies, and it is ``pushed'' into the $\omega <
\omega_R/2$ regime, where Beliaev damping of the polariton is not
allowed.  Here, in the spectra associated with $y/y_\text{c} =0.9$
(dashed-double-dotted brown line) and $y/y_\text{c} =0.95$
(long-dashed magenta line), there is a strong sharp peak below
$\omega_R/2$ which has a linewidth determined by $\kappa$ and the
coupling strength $y$. The tail of these peaks leaks into the $\omega
> \omega_R/2$ region. The main feature in
Fig.~\ref{fig:Correlation_c_mode}b is the spectacular hole burning
effect at $\omega=\omega_R/2$ for $y/y_\text{c} =0.9$
(dashed-double-dotted brown line) which originates from the
non-trivially structured coupling density function of the Beliaev
damping process.

In Fig.~\ref{fig:Correlation_c_mode}c, one can see the effect of a
much larger cavity decay parameter $\kappa$. The fine structure of the
spectra $C_c(\omega)$ near the frequency $\omega_R/2$ is washed out.
Only a small dip can be observed on the curve for $y/y_\text{c} =0.95$
(long-dashed magenta line). The higher $\kappa$ amounts to larger
fluctuations that can excite efficiently the mode $\hat c$, thus the
peak of the correlation function gets increased. Meanwhile the peak is
broadened significantly in the $\omega < \omega_R/2$ regime. 

Figures~\ref{fig:Correlation_a_mode}a-c show the corresponding
spectrum of fluctuations in the photon mode $a$ which can be directly
measured by photodetectors. Fig.~\ref{fig:Correlation_a_mode}a
presents the full correlation function, including the cavity resonance
peak at $\Delta_C = 100\omega_R$. This Lorentzian peak
(dashed-double-dotted red line) is the same for all coupling
strengts. In contrast, the low frequency part of the spectrum
(expanded in the inset) shows the hybridization of the cavity mode $a$
with the atomic excitation mode $c$. For nonzero coupling ($y>0$)
these peaks corresponds to the ones observed in
Fig.~\ref{fig:Correlation_c_mode}a. In
Fig.~\ref{fig:Correlation_a_mode}b-c we present only the interesting
low frequency spectrum. The features observed at the atomic mode
(Figs.~\ref{fig:Correlation_c_mode}b-c) appear similarly in the
photonic spectrum, which facilitate direct observation of the effect.

\begin{figure}
\includegraphics[width=0.75\linewidth]{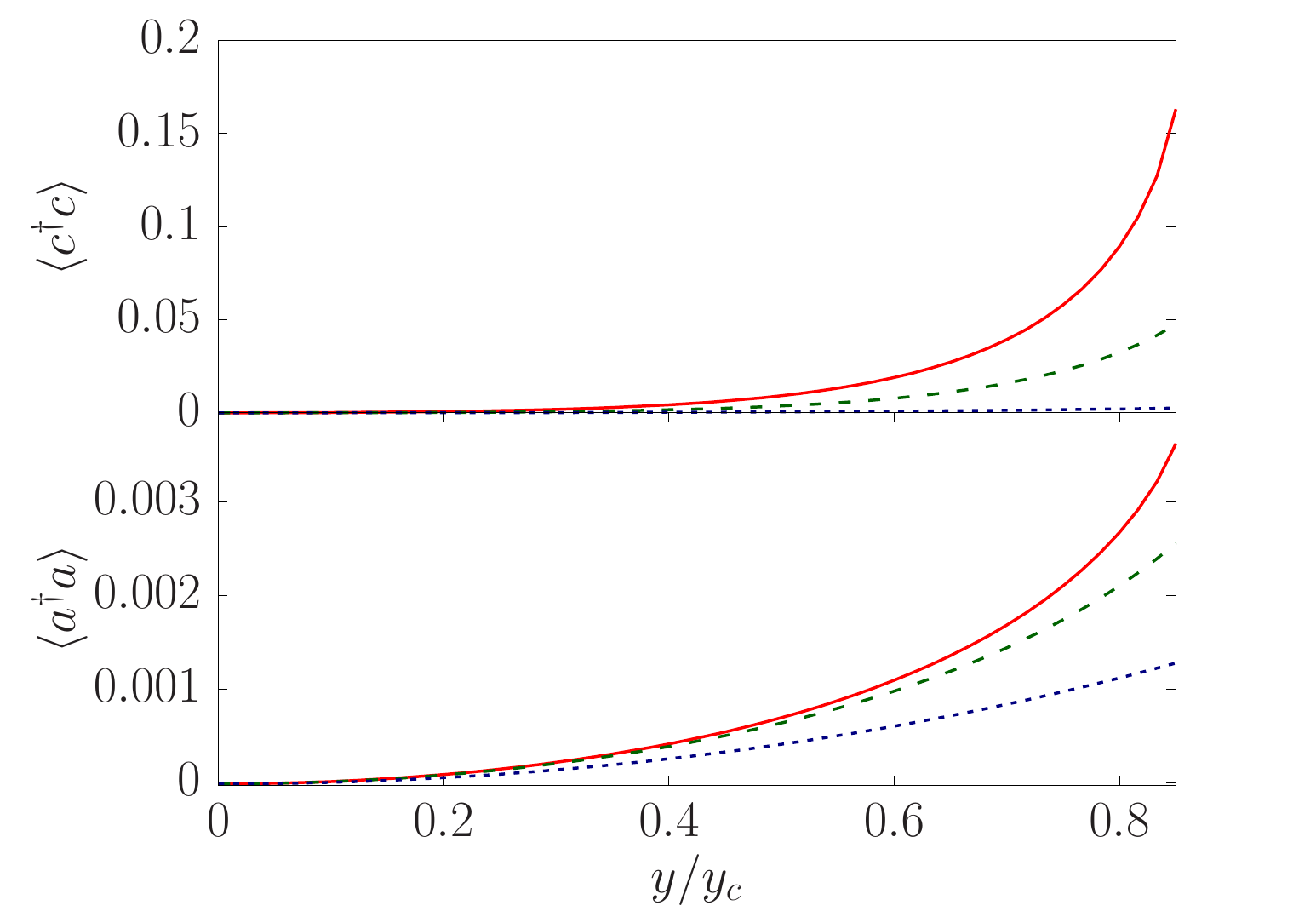}
\caption{Steady-state populations in the atomic mode (top panel) and
  in the cavity field mode (bottom panel) for three different Beliaev
  scattering rates: $\dsgamma = 0.01$ (solid red), $0.1$
  (dashed green) and $1\,\omega_R$ (dotted blue). }
\label{fig:populations}
\end{figure}

Finally, we calculate the excitation numbers in the steady-state.
They are given by the equal-time correlations, i.e. $\langle
c^\dagger(0) c(0) \rangle = (C_c(t=0) - 1)/2$, which is obtained by
the integral of the spectrum \mbox{$C_{c}(t=0) = \int
  \frac{\ud\omega}{2\pi} C_{c}(\omega)$}. In
Fig.~\ref{fig:populations} we plot the excitation number of mode $c$
(top panel) and of the cavity mode $a$ (bottom panel) as a function of
the coupling $y$. Higher $\dsgamma$ results in lower
steady-state populations.1 Although the spectrum of fluctuations were
found to be significantly structured functions due to the Beliaev
damping effect, the total number of fluctuations follow a simple
monotonic dependence as a function of the control parameter $y$. Due
to the Dicke-type phase transition in the system, the excitation
numbers in each mode diverge at the critical point $y_c$. The critical
exponent, however, cannot be changed by the Beliaev damping, since its
coupling-density function vanishes below $\omega_R/2$, while the system
reaches criticality, when the soft mode frequency tends to zero. 
Therefore, the critical behaviour is determined by the cavity decay process as in
Ref.~\cite{Nagy2011Critical}.

\section{Conclusions}

By using the Keldysh-type Green's function approach, we revisited the Beliaev damping process acting on the quasi-particle excitation of a condensate coupled to a photonic mode of an optical resonator. We showed that the effect of phonons in a four-particle scattering process can be mapped to the problem of linear coupling to a reservoir composed of bosonic modes.  The presented general approach confirms a previous result that the Beliaev scattering effect has a significant influence on the spectrum of fluctuations, giving rise to a peak at a certain mixing ratio of the quasi-particle excitation and photon mode in the polariton. The coupling density function characterizing the effective reservoir is a strongly patterned spectral function which has a non-analytic point at half the recoil frequency. As a consequence, the measurable spectrum of fluctuations of both the quasi-particle and the photonic modes clearly reflect non-Markovian dynamics, which is beyond the enhanced damping effect  \cite{Konya2014Photonic,Konya2014Damping}. The spectrum of fluctuations exhibiting a hole burning feature reveal an interesting interplay between the photon loss and the Beliaev scattering processes. On the other hand, the singularity of the Beliaev process occurs in a frequency range away form the critical point where the soft polariton mode frequency vanishes. Therefore the exponent of the self-organization criticality is not affected  by the Beliaev damping process.

\section*{Acknowledgements}

This work was supported by the National Research, Development and Innovation Office of Hungary (NKFIH) within the Quantum Technology National Excellence Program  (Project No. 2017-1.2.1-NKP-2017-00001) and by Grant No. K115624. D. Nagy was supported  by the J\'anos Bolyai Fellowship of the Hungarian Academy of Sciences.

\appendix

\section{Perturbation theory in the Keldysh formalism}
\label{app:pertkel}

In Sec.~\ref{sec:keldysh}, the main goal is to arrive to an effective model for the modes $a$ and $c$ by integrating out the phonon modes $h_p$. In other words, we introduce a new dissipation channel, namely the damping of the cosine mode to other phonons of the system. In this appendix, we explicitly perform this integration at one-loop level. We start from the definition of the Green's function of the cosine mode, given by Eq.~\eqref{eq:intgrmat}. We refer to the components of the matrix in the following way:\begin{equation}
iG_{c,\alpha\beta}(t,t')=\langle c^\alpha(t)\,c^{\beta *}(t')\rangle=\int \cl{D} \left[ a,c,h \right] c^{\alpha} (t) \, c^{\beta \, *} (t') \, e^{i S},
\label{eq:Grfunwav}
\end{equation}
where the Greek indices take values from the set $\lbrace cl, q\rbrace$, or $\lbrace 1,2\rbrace$, respectively. The path integral $\int\cl{D} \left[ a,c,h \right]\ldots$ is over both classical and quantum components of $a$, $c$, and $h_p$, and in the case of $h_p$ also over all wavenumbers $p\in\cl{P}^{(L)}$. The action $S$ is given by Eq.~\eqref{eq:Keldactf}. Formally, one can recast the integrals to
\begin{multline}
\int \cl{D} \left[ a,c,h \right] c^{\alpha} (t) \, c^{\beta \, *} (t') \, e^{i S}\\
=\int \cl{D} \left[ a,c\right] c^{\alpha} (t) \, c^{\beta \, *} (t') \, e^{i (S_a+S_c+S_{ac})}\int \cl{D} \left[h\right]e^{i S_h} e^{iS_{ch}}.
\end{multline}
The second factor in the second line depends on the variables $c^{cl}$ and $c^{q}$, c.f., Eq.~\eqref{eq:sint}. Thus, by performing the integral over the $h_p$ fields we end up with something depending on the $c$ fields in a nontrivial way. A Gaussian approximation of this function is achieved by expanding $e^{iS_{ch}}$ up to second order. The integrals over $h_p$ can be performed with the help of the Wick theorem. The Feynman graphs are given in Fig.~\ref{fig:self_energ}, and the approximation corresponds to one-loop level. In the end, we define $S_{\Sigma}$ by
\begin{equation}
\int \cl{D} \left[h\right]e^{i S_h} e^{iS_{ch}}\approx \int \cl{D} \left[h\right]e^{i S_h} \bigg(1-\frac{S_{ch}^2}{2}\bigg) = 
1+iS_{\Sigma}\approx e^{i S_{\Sigma}},
\label{eq:integration}
\end{equation}
where
\begin{equation}
S_{\Sigma}\\
= -\int_{-\infty}^{+\infty} \frac{\ud\omega}{2\pi} \, 
\begin{pmatrix} 
c^{cl \, *} (\omega) & c^{q \, *} (\omega)
\end{pmatrix}
\boldsymbol{\Sigma}_c (\omega) 
\begin{pmatrix}
c^{cl} (\omega) \\
c^{q} (\omega) 
\end{pmatrix} .
\label{eq:effaction}
\end{equation}
As a result, we end up with the effective action of the photon-polariton dynamics,
\begin{equation}
S_{\text{eff}}=S_a + \widetilde{S}_c + S_{ac}\,,
\label{eq:effaction2}
\end{equation}
where
\begin{equation}
  \widetilde{S}_c = S_c+S_\Sigma\,.
\end{equation}
Here, $S_a$, $S_c$ and $S_{ac}$ are defined by Eqs.~\eqref{eq:Sa},~\eqref{eq:Sc}, and~\eqref{eq:Sac}, respectively. Equivalently, we can define an intermediate Green's function, according to Eq.~\eqref{eq:internalG}, with which
\begin{equation}
\widetilde{S}_c = \int_{-\infty}^{+\infty}\frac{\ud\omega}{2\pi} \, 
\begin{pmatrix} 
c^{cl \, *} (\omega) & c^{q \, *} (\omega)
\end{pmatrix} 
\widetilde{\boldsymbol{G}}_c^{-1}(\omega)
\begin{pmatrix}
c^{cl} (\omega) \\
c^{q} (\omega) 
\end{pmatrix} .
\label{eq:internalG2}
\end{equation}

\bibliography{szirmai}

\end{document}